# Speaker recognition improvement using blind inversion of distortions[1]

*Marcos Faúndez-Zanuy (\*), Jordi Solé-Casals (\*\*)*

(\*) Escola Universitària Politècnica de Mataró, UPC
(\*\*) Signal Processing Group, Universitat de VIC
BARCELONA (SPAIN)
`faundez@eupmt.es, jordi.sole@uvic.es`

## Abstract

In this paper we propose the inversion of nonlinear distortions in order to improve the recognition rates of a speaker recognizer system. We study the effect of saturations on the test signals, trying to take into account real situations where the training material has been recorded in a controlled situation but the testing signals present some mismatch with the input signal level (saturations). The experimental results shows that a combination of data fusion with and without nonlinear distortion compensation can improve the recognition rates with saturated test sentences from 80% to 88.57%, while the results with clean speech (without saturation) is 87.76% for one microphone.

## 1. Introduction

This paper proposes a non-linear channel distortion estimation and compensation in order to improve the recognition rates of a speaker recognizer. Mainly it is studied the effect of a saturation on the test signals and the compensation of this non-linear perturbation. This paper is organized as follows. Section 2 describes the Wiener model, its parameterization, and obtains the cost function based on statistical independence. Section 3 summarizes the speaker recognition application. Section 4 deals the experiments using the blind inversion in conjunction with the speaker recognition application.

## 2. Non-parametric approach to blind deconvolution of nonlinear channels

When linear models fail, nonlinear models appear to be powerful tools for modeling practical situations. Many researches have been done in the identification and/or the inversion of nonlinear systems. These works assume that both the input and the output of the distortion are available [1]; they are based on higher-order input/output cross-correlation [2], bispectrum estimation [3, 4] or on the application of the Bussgang and Prices theorems [5, 6] for nonlinear systems with Gaussian inputs. However, in a real world situations, one often does not have access to the distortion input. In this case, blind identification of the nonlinearity becomes the only way to solve the problem. This paper is concerned by a particular class of nonlinear systems, composed by a linear filter followed by a memoryless nonlinear distortion (figure 1, up). This class of nonlinear systems, also known as a Wiener system, is a nice and mathematically attracting model, but also a realistic model used in various areas. We use a fully blind inversion method inspired of recent advances in source separation of nonlinear mixtures [7]. Although deconvolution can be viewed as a single input/single output (SISO) source separation problem in convolutive mixtures (which are consequently not cited in this paper), the current approach is actually very different. It is mainly based on equivalence between instantaneous postnonlinear mixtures and Wiener systems, provided a well-suited parameterization.

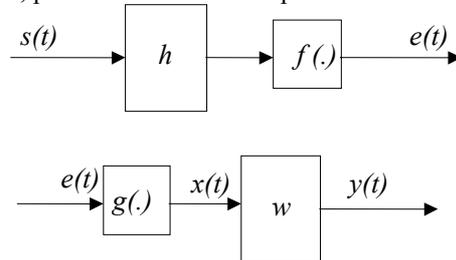

Figure 1: The unknown nonlinear convolution system (up) and the proposed inversion structure (down).

### 2.1. Model and assumptions

We suppose that the input of the system $S=\{s(t)\}$ is an unknown non-Gaussian independent and identically distributed (i.i.d.) process, and that subsystems $h, f$ are a linear filter and a memoryless nonlinear function, respectively, both unknown and invertible. We would like to estimate $s(t)$ by only observing the system output. This implies the blind estimation of the inverse structure (figure 1, down), composed of similar subsystems: a memoryless nonlinear function $g$ followed by a linear filter $w$. Such a system is known as a Hammerstein system. Let **s** and **e** be the vectors of infinite dimension, whose $t$-th entries are $s(t)$ or $e(t)$,

---

[1] This work has been supported by CICYT TIC2000-1669-C04-02, European COST action 277, and University of Vic (R0912)

respectively. The unknown input-output transfer can be written as:

$$e = f(\mathbf{H}s) \quad (1)$$

where:

$$\mathbf{H} = \begin{pmatrix} \cdots & \cdots & \cdots & \cdots & \cdots \\ \cdots & h(t+1) & h(t) & h(t-1) & \cdots \\ \cdots & h(t+2) & h(t+1) & h(t) & \cdots \\ \cdots & \cdots & \cdots & \cdots & \cdots \end{pmatrix} \quad (2)$$

is an infinite dimension Toeplitz matrix which represents the action of the filter $h$ to the signal $s(t)$. The matrix $\mathbf{H}$ is non-singular provided that the filter $h$ is invertible, i.e. satisfies $h^{-1}*h(t) = h*h^{-1}(t) = \delta(t)$, where $\delta(t)$ is the Dirac impulse. The infinite dimension of vectors and matrix is due to the lack of assumption on the filter order. If the filter $h$ is a finite impulse response (FIR) filter of order $N_h$, the matrix dimension can be reduced to the size $N_h$. Practically, because infinite-dimension equations are not tractable, we have to choose a pertinent (finite) value for $N_h$.

Equation (1) corresponds to a post-nonlinear (pnl) model [8]. This model has been recently studied in nonlinear source separation, but only for a finite dimensional case. In fact, with the above parameterization, the i.i.d. nature of $s(t)$ implies the spatial independence of the components of the infinite vector $s$. Similarly, the output of the inversion structure can be written $y = \mathbf{W}x$ with $x(t) = g(e(t))$. Following [8, 9] the inverse system $(g, w)$ can be estimated by minimizing the output mutual information, i.e. spatial independence of $y$ which is equivalent to the i.i.d. nature of $y(t)$.

### 2.2. Cost function

The mutual information of a random vector of dimension $n$, defined by

$$I(\mathbf{z}) = \sum_{i=1}^{n} H(z_i) - H(z_i, z_2, ..., z_n) \quad (3)$$

can be extended to a vector of infinite dimension, using the notion of *entropy rates* of stationary stochastic processes [10]

$$\lim_{T \to \infty} \frac{1}{2T+1} \left\{ \sum_{t=-T}^{T} H(z(t)) - H(z(-T), ..., z(T)) \right\}$$

$$I(z) = H(z(\tau)) - H(Z) \quad (4)$$

where $\tau$ is arbitrary due to the stationarity assumption. We can notice that $I(Z)$ is always positive and vanishes iff $z(t)$ is i.i.d. Since $S$ is stationary, and $h$ and $w$ are time-invariant filters, then $Y$ is stationary too, and $I(Y)$ is defined by

$$I(Y) = H(y(\tau)) - H(Y) \quad (5)$$

Using the Lemma 1 of [9], the last right term of equation (5) becomes

$$H(Y) = H(X) + \frac{1}{2\pi} \int_0^{2\pi} \log \left| \sum_{t=-\infty}^{+\infty} w(t) e^{-jt\theta} \right| d\theta \quad (6)$$

Moreover, using $x(t) = g(e(t))$ and the stationarity of $E = \{e(t)\}$:

$$H(X) = \lim_{T \to \infty} \frac{1}{2T+1} \Big\{ H(e(-T), ..., e(T)) + \sum_{t=-T}^{T} E\Big[\log\ g'(e(t))\Big] \Big\} \quad (7)$$

$$= H[E] + E\Big[\log\ g'(e(t))\Big]$$

Combining (6) and (7) in (5) leads finally to:

$$I(Y) = H(y(\tau)) - \frac{1}{2\pi} \int_0^{2\pi} \log \left| \sum_{t=-\infty}^{+\infty} w(t) e^{-jt\theta} \right| d\theta$$

$$-E\Big[\log\ g'(e(\tau))\Big] - H[E]$$

### 3. Speaker recognition

One of the main sources of degradation of the speaker recognition rates is the mismatch between training and testing conditions. For instance, in [11] we evaluated the relevance of different training and testing languages, and in [12] we also studied other mismatch, such as the use of different microphones.

In this paper, we study a different source of degradation: different input level signals in training and testing. Mainly we consider the effect of a saturation. We try to emulate a real scenario where a person speaks too close to the microphone or to loud, producing a saturated signal. Taking into account that the perturbations are more damaging when they are present just during training or testing but not in both situations, we have used a clean database and artificially produced a saturation in the test signals. Although it would be desirable to use a "real" saturated database, we don't have this kind of database, and the simulation let us more control about "how the algorithm is performing".

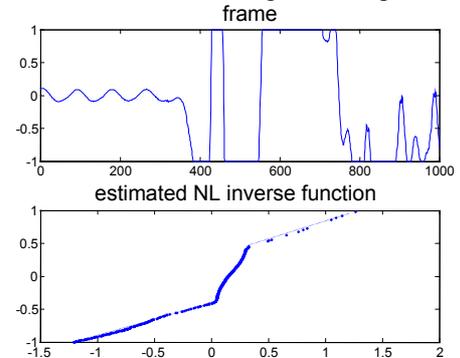

Figure 2. saturated frame and the estimated channel function.

Anyway, we have used a real saturated speech sentence in order to estimate the nonlinear distortion using the algorithm described in section 2 and the results have been successful. Figure 2 shows a real saturated speech frame and the corresponding estimate of the NL perturbation.

### 3.1. Database

For our experiments we have used a subcorpora of the Gaudi database, that follows the design of [13]. It consists on 49 speakers acquired with a simultaneous stereo recording with two different microphones (AKG C-420 and SONY ECM66B). The speech is in wav format at fs=16 kHz, 16 bit/sample and the bandwidth is 8 kHz.

We have applied the potsband routine that can be downloaded from:
http://www.ee.ic.ac.uk/hp/staff/dmb/voicebox/voicebox.html, in order to obtain narrow-band signals. This function meets the specifications of G.151 for any sampling frequency.

The speech signals are pre-emphasized by a first order filter whose transfer function is $H(z)=1-0.95z^{-1}$. A 30 ms Hamming window is used, and the overlapping between adjacent frames is 2/3. One minute of read text is used for training, and 5 sentences for testing (each sentence is about two seconds long).

### 3.2. Speaker recognition algorithm

We have chosen a second-order based measure for the recognition of a speaker.

In the training phase, we compute for each speaker empirical covariance matrices (CM) based on feature vectors extracted from overlapped short time segments of the speech signals, i.e., $C_j = \hat{E}[x_n x_n^T]$, where $\hat{E}$ denotes estimate of the mean and $x_n$ represents the features vector for frame $n$. As features representing short time spectra we use mel-frequency cepstral coefficients.

In the speaker-recognition system, the trained covariance matrices (CM) for each speaker are compared to an estimate of the covariance matrix obtained from a test sequence from a speaker. An arithmetic-harmonic sphericity measure is used in order to compare the matrices [14]:

$$d = \log\left(\mathrm{tr}(C_{test}C_j^{-1})\,\mathrm{tr}(C_j C_{test}^{-1})\right) - 2\log(l),$$

where $\mathrm{tr}(\cdot)$ denotes the trace operator, $l$ is the dimension of the feature vector, $C_{test}$ and $C_j$ is the covariance estimate from the test speaker and speaker model $j$, respectively.

## 4. Experiments and conclusions

Using the database described in section 3, we have artificially generated a test signal database, using the following procedure:

All the test signals are normalized to achieve unitary maximum amplitude.
- A saturated database has been artificially created using the following equation:
$x' = \tanh(kx)$, where $k$ is a positive constant.

The training set remains the same, so no saturation is added.

In order to show the improvement due to the compensation method, figure 3 shows one frame that has been artificially saturated with a dramatic value ($k$=10), the original, and the recovered frame applying the blind inversion of the distortion.

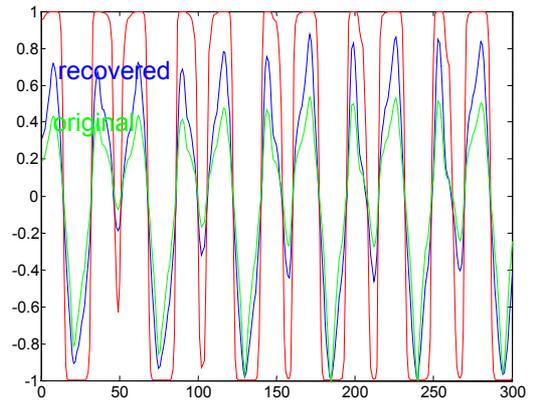

Figure 3. Example of original, saturated, and recovered frame using the proposed procedure.

Using the original (clean) and artificially generated database (saturated) we have evaluated the identification rates.

For the saturated test sentences scenario, we have estimated one different channel model for each test sentence, applying the method described in section 2. This a way to manage real situations where the possible amount of saturation is not known in advance and must be estimated for each particular test sentence.

| Combination | | Recognition rate |
|---|---|---|
| 1 (AKG+NL compensation) | | 83.67 % |
| 2 (AKG) | | 82.04 % |
| 3 (SONY+NL compensation) | | 80.82 % |
| 4 (SONY) | | 80 % |
| fusion 1&2 | Arithmetic mean | 84.9 % |
| | Geometric mean | 84.9 % |
| fusion 1&3 | Arithmetic mean | 88.57% |
| | Geometric mean | 88.57% |
| fusion 2&4 | Arithmetic mean | 87.35% |
| | Geometric mean | 87.35 % |
| fusion 1&2&3&4 | Arithmetic mean | 88.16 % |
| | Geometric mean | 87.76 % |

Table 1 Results for several fusion criteria, shown in figure 4.

In order to improve the results a opinion fusion is done, using the scheme shown in figure 4. Thus, we present

the results in three different combination scenarios for speaker recognition:
- just one opinion (1 or 2 or 3 or 4)
- To use the fusion of two opinions (1&2 or 2&3).
- The combination of the four available opinions.

Table 1 shows the results for *k*=2 in all these possible scenarios using two different combinations rules (arithmetic and geometric mean) [15].

Main conclusions are:
- The use of the NL compensation improves the results obtained with the same conditions without this compensation block.
- The combination between different classifiers improves the results. These results can be even more improved using a weighted sum (more than 90% identification rate)
- We think that using a more suitable parameterization, the improvements would be higher.

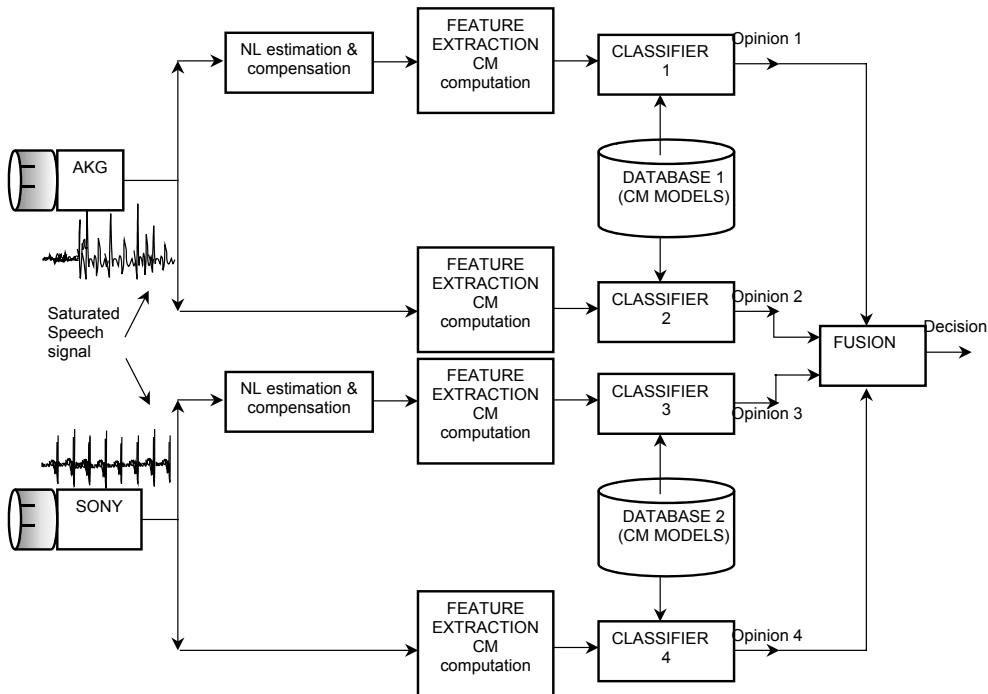

Figure 4. General Scheme of the recognition system